\documentclass[letterpaper]{article} 
\usepackage{aaai2026}  
\usepackage{times}  
\usepackage{helvet}  
\usepackage{courier}  
\usepackage[hyphens]{url}  
\usepackage{graphicx} 
\usepackage{natbib}  
\usepackage{caption} 
\usepackage{algorithm}
\usepackage{algorithmic}

\usepackage{booktabs}
\usepackage{tabularx}
\usepackage{multirow}
\usepackage{array}
\usepackage{subfigure}
\usepackage{amsmath}
\usepackage{arydshln}
\usepackage{xcolor}
\usepackage[frozencache,cachedir=minted-cache]{minted}
\usepackage[skins,breakable]{tcolorbox}
\newtcolorbox{PromptBlock}{
    breakable,
    colback=gray!20,
    colframe=black,
    arc=3mm,
    title={},
    toptitle=0mm,
    bottomtitle=0mm,
    colbacktitle=gray!20,
    coltitle=black,
}

\usepackage{newfloat}
\usepackage{listings}
\DeclareCaptionStyle{ruled}{labelfont=normalfont,labelsep=colon,strut=off} 
\floatstyle{ruled}
\newfloat{listing}{tb}{lst}{}
\floatname{listing}{Listing}
\nocopyright 

\title{ContextASR-Bench: A Massive Contextual Speech Recognition Benchmark}
\author{
    He Wang,
    Linhan Ma,
    Dake Guo,
    Xiong Wang,
    Lei Xie,
    \\
    Jin Xu\thanks{Corresponding Author},
    Junyang Lin
}
\affiliations{
    Alibaba Group, Beijing, China \\
    
    \texttt{\{wanghe.wh2001,jxu3425\}@alibaba-inc.com}
}

\usepackage{bibentry}

\begin{document}

\maketitle

\begin{abstract}
Automatic Speech Recognition (ASR) has been extensively investigated, yet prior benchmarks have largely focused on assessing the acoustic robustness of ASR models, leaving evaluations of their linguistic capabilities relatively underexplored. This largely stems from the limited parameter sizes and training corpora of conventional ASR models, leaving them with insufficient world knowledge, which is crucial for accurately recognizing named entities across diverse domains. For instance, drug and treatment names in medicine or specialized technical terms in engineering. Recent breakthroughs in Large Language Models (LLMs) and corresponding Large Audio Language Models (LALMs) have markedly enhanced the visibility of advanced context modeling and general artificial intelligence capabilities. Leveraging LLMs, we envision a unified system capable of robust speech recognition across diverse real-world domains, yet existing benchmarks are inadequate for evaluating this objective. To address this gap, we propose ContextASR-Bench: a comprehensive, large-scale benchmark designed to assess the linguistic competence of ASR systems using corpora that feature numerous named entities across multiple domains. It encompasses up to 40,000 data entries with more than 300,000 named entities across over 10 domains. Beyond the audio and its transcription, each sample provides the domain it belongs to and a list of named entities it contains, which are referred to as the context. Based on this, we introduce three evaluation modes to assess how effectively models can exploit such context to improve ASR accuracy. Extensive evaluation on ContextASR-Bench highlights that LALMs outperform conventional ASR models by a large margin thanks to the strong world knowledge and context modeling of LLMs, yet there remains ample room for further improvement. The dataset and evaluation code have been released at \color{blue}{https://github.com/MrSupW/ContextASR-Bench}.
\end{abstract}


\section{Introduction}
Automatic Speech Recognition (ASR) transcends a mere mapping task between speech and text modalities. 
Human comprehension of spoken content necessitates the integration of extensive world knowledge acquired through learning processes and a nuanced understanding of the contextual elements inherent in auditory input.
For example, even a Ph.D.-level computer scientist might struggle to accurately transcribe dialogues within medical contexts due to insufficient related domain knowledge. 
Likewise, ASR systems often encounter similar problems in real-world applications: even when the acoustic environment is excellent, virtually free of noise or interference, the lack of relevant linguistic knowledge can cause the model to omit or misrecognize words, typically producing homophonic substitution errors where the sounds are correct but the words are wrong.
Conventional ASR models~\citep{jointCTC, conformer, transducer, paraformer, SenseVoice, whisper} have long been constrained by their limited capacity to integrate comprehensive world knowledge and contextual nuances, typically only excelling in specific domains or casual conversational contexts. 
Therefore, additional effort is often required to adapt an ASR model to the desired deployment scenario or domain, and a common approach is to fine-tune the model using data relevant to the target domain~\citep{domain_adaptation_filtering, domain_adaptation_synthetic}.
Statistical N-gram language models~\citep{improving_mandarin, NGPU-LM} and Neural Network-based language models~\citep{shallow_fusion, graph_neural_networks} are also commonly used to provide additional contextual information.
However, due to limitations primarily arising from data volume and model size, these methods have not actually achieved building an ASR model that is suitable for all domains.

Recent advancements in general artificial intelligence, particularly reflected through the development of Large Language Models (LLMs)~\citep{qwen3, gpt4, deepseekr1} and Large Audio Language Models (LALMs)~\citep{qwen2audio, qwen25omni, kimi-audio}, which typically consist of an audio encoder and an LLM backbone, have demonstrated a substantial capability in encoding comprehensive world knowledge and performing complex reasoning tasks. 
By leveraging the knowledge that LLMs absorb from massive text training corpora, we can envision that a single LALM can perform speech recognition robustly across diverse domains and can be further tailored to a specific field with minimal effort, for example, by simply adding domain cues within the user prompt.
However, LALMs do not exhibit overwhelming advantages over conventional ASR models in speech recognition as they do in other tasks (see Section~\ref{fig:existing_bench} for details), which is quite counterintuitive.
Given that the current ASR benchmarks~\citep{aishell1, librispeech} typically have short utterances with narrow domains and casual conversational corpora, they do not effectively showcase how the powerful contextual modeling capabilities and extensive world knowledge across almost all domains of LLMs can enhance the ASR performance. 
Therefore, there is an urgent need for a new ASR benchmark with longer speech recordings and more challenging multi-domain corpora, including technical terms and named entities, to evaluate the upper limits of LLM-based ASR systems.

In this paper, we propose \textbf{\textit{ContextASR-Bench}}, a comprehensive benchmark for contextual speech recognition, with over 40,000 test pairs, aiming to evaluate the linguistic abilities of ASR systems. 
A broad spectrum of text corpora is adopted, encompassing various domains and incorporating rich named entities. 
Subsequently, these corpora served as seeds for strong LLMs (e.g., DeepSeek-R1) to generate colloquial text along with the domain label it belongs to and a list of named entities it contains. 
To focus on evaluating the linguistic competence of ASR systems while eliminating potential acoustic confounds, we construct a text-to-speech (TTS) synthesis pipeline that employs strong Zero-Shot TTS models~\citep{cosyvoice2, xtts} to convert generated text into speech.
To enhance the voice diversity, the speaker timbre of each speech is randomly selected from an inventory of over 20,000 reference speakers sourced from open-source speech datasets~\citep{wenetspeech4tts, emilia}.
For reliability assurance, we employ two ASR systems to transcribe the synthetic speech for cross-validation. 
The Phoneme Error Rate (PER) between the original text and the transcription is calculated, and only synthetic speech with a PER below a predefined threshold will be retained.

ContextASR-Bench includes two test sets: \textbf{\textit{ContextASR-Speech}} set and \textbf{\textit{ContextASR-Dialogue}} set. 
The former uses open-source Named Entity Recognition (NER) datasets~\citep{ner_cmeee, ner_cluener, ner_mmc, ner_dlner} as the seeds for DeepSeek-R1~\citep{deepseekr1} to generate colloquial text, and then synthesizes single-speaker speech.
The latter leverages curated movie information crawled from the internet as seeds to generate dialogue text discussing the plot and characters, featuring multi-speaker dialogue speech.
These sets substantially improve the corpus diversity and facilitate the assessment of model capabilities in multi-speaker speech recognition. 
Detailed statistics of these two test sets can be found in Table~\ref{tab:detailed_statistics}.
The evaluation within our benchmark is divided into three modes: \textbf{\textit{Contextless}} setting, \textbf{\textit{Coarse-grained Context}} setting, and \textbf{\textit{Fine-grained Context}} setting. 
The first setting directly assesses the models' speech recognition abilities without any additional contextual input. 
The second setting provides coarse-grained contextual cues, such as the domain label of the utterance or the movie title around which the conversation revolves, to evaluate the ASR system’s ability to leverage this rough context to improve speech recognition accuracy.
The third setting examines models' proficiency in comprehending fine-grained contexts mentioned in the auditory input, such as technical terms, named entities, or personal names.
For evaluation, we use Named Entity WER (NE-WER) and Named Entity False Negative Rate (NE-FNR) metrics to assess models' accuracy in recognizing named entities, mainly constituting the knowledge across various domains. 
NE-WER is calculated between the words of extracted entities in transcriptions.
NE-FNR is the ratio of the number of entities in the transcription that are not accurately recognized to the total number of entities.
\begin{table}[tp]
\centering
\fontsize{9}{11}\selectfont{
    \begin{tabularx}{\linewidth}{ccccc} 
    \toprule
    Subset & Lang & Utterance & Duration~(h) & Entities \\ 
    \midrule
    ContextASR-     & EN & 15,326 & 187.98 & 116,167 \\
    Speech          & ZH & 15,498 & 197.64 & 97,703  \\ \midrule
    ContextASR-     & EN & 5,273 & 221.86 & 58,741  \\
    Dialogue        & ZH & 5,232 & 230.39 & 50,250  \\ \bottomrule
    \end{tabularx}
}
    \caption{
        Detailed statistics on ContextASR-Bench, comprising two parts: ContextASR-Speech and ContextASR-Dialogue, each containing Mandarin (ZH) and English (EN) databases. ``Utterance'' refers to the number of data entries, ``Duration'' refers to the total duration of speech data, and ``Entities'' refers to the number of named entities included.
    }
\label{tab:detailed_statistics}
\end{table}

\begin{figure*}[ht]
    \centering
    \includegraphics[width=\textwidth]{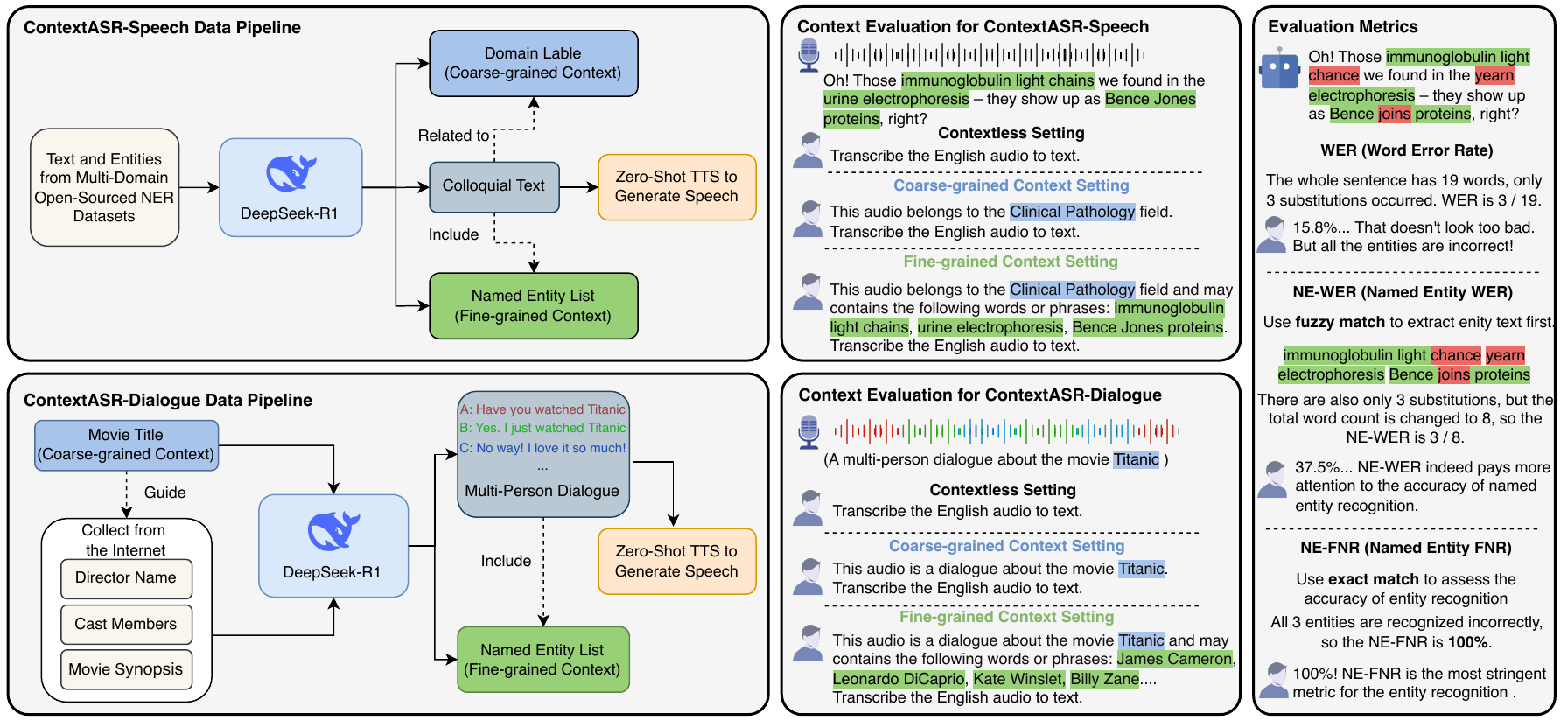}
    \caption{
        An overview of our proposed ContextASR-Bench, comprising ContextASR-Speech and ContextASR-Dialogue. The left part shows the data pipeline for these two test sets. Both use DeepSeek-R1 to generate entity-rich corpora, which are then synthesized into speech using Zero-Shot TTS. Each entry in both sets follows the same data structure: $<$\textbf{Audio, Text, Coarse-grained Context, Fine-grained Context}$>$. The middle part presents three contextual evaluation settings. The contextless setting can be used for evaluating any ASR systems, while the other two assess LALMs' context comprehension capacity through different granularity context information within the prompt. The right part introduces three evaluation metrics used in ContextASR-Bench, where NE-WER and NE-FNR focus more on the accuracy of named entity recognition compared to WER. 
    }
    \label{fig:main_figure}
\end{figure*}

We present a comprehensive evaluation of both conventional ASR models and LALMs, which shows conventional ASR systems without LLMs struggle significantly in ContextASR-Bench compared to LALMs. 
This demonstrates the importance of the world knowledge possessed by LLMs for speech recognition tasks in specialized domains. 
The contribution of the paper is summarized as follows:
\begin{itemize}
    \item We open-source \textit{\textbf{ContextASR-Bench}}, the first massive contextual ASR benchmark comprising over 40,000 data entries, focused on assessing the linguistic capabilities of ASR models using corpora with over 300,000 technical terms and named entities across more than 10 domains in both English and Mandarin. 
    \item We perform a comprehensive evaluation of open-source ASR models and LALMs on ContextASR-Bench. Experimental results show that LALMs achieve a large performance lead over conventional ASR systems due to their extensive world knowledge of LLMs, particularly in recognizing named entities across multiple domains. Despite this advantage, current LALMs still struggle in contextual ASR, indicating ample room for improvement.
\end{itemize}

\section{Methods}
Obtaining large-scale, entity-rich speech-text paired data from real-world scenarios poses significant challenges, particularly in managing thematic distribution, diversity levels, and entity density within naturally occurring data. 
To address these obstacles, we design an innovative data pipeline that integrates LLM-driven entity-rich text generation with Zero-Shot TTS systems. 
This section details the architectural components of this data pipeline and presents the evaluation framework of ContextASR-Bench.

\subsection{Entity-rich Corpora Generation}
\label{sec:entity-rich_corpara_generation}
To efficiently evaluate the recognition accuracy of ASR systems for specialized domain terms or named entities, the primary task involves preparing entity-rich corpora to serve as text contents for subsequent speech generation in ContextASR-Bench. 
LLMs~\citep{qwen3, gpt4, deepseekr1}, trained on vast textual datasets, demonstrate exceptional world knowledge comprehension far beyond that of any individual human.
It also includes the understanding of technical terms or named entities in various fields. 
This makes LLMs particularly suitable for generating multi-domain entity-rich corpora. 
Therefore, we design an approach for constructing entity-rich corpus data based on LLM by incorporating seeds into LLM prompts to ensure the diversity and controllability of the generated results, as shown in the left part of Figure~\ref{fig:main_figure}.

ContextASR-Speech set aims to evaluate the performance of ASR systems in recognizing technical terms or named entities across various domains.
Firstly, we collect publicly available open-source text NER datasets. We include details in Appendix A.
While these datasets provide annotated texts with domain-specific entities across multiple fields, they predominantly contain formal written language from web sources or publications, significantly differing from colloquial speech patterns. 
Specifically, we found that the variable text lengths (ranging from a few words to thousands) and sparse entity distribution in NER datasets render them inappropriate for contextual ASR evaluation, but on the other hand, their extensive domain coverage makes them ideal seeds for the LLMs to generate entity-rich text, so it is necessary and feasible to use LLM for transforming the original NER data entry into colloquial text with a suitable text length and named entity density. 
For the choice of LLM, we use the open-source DeepSeek-R1~\citep{deepseekr1}, which demonstrates strong writing and instruction following capabilities in corresponding text benchmarks~\citep{alpacaeval, arenahard_bench, MMLU, IFEval}. 
In addition, we establish two key requirements in the prompt for DeepSeek-R1: 1) Generate colloquially styled texts based on the raw NER text and annotated entities within it, 2) Expand the entities intentionally to raise the entity density as the fine-grained context, and 3) Summarize the domain label the LLM generated colloquial text and entity list related to as the coarse-grained context. 
Detailed prompt content can be found in Appendix C1. 



ContextASR-Dialogue set focuses on the personal name recognition accuracy of ASR systems and the robustness of the multi-speaker dialogue format audio. 
As we know, movies serve as artistic carriers of characters and stories, and when people discuss a movie, the names of actors or characters are frequently mentioned.
Therefore, we select multi-speaker discussions on a certain movie as the testing scenario for ContextASR-Dialogue. 
Based on recent popular movie titles, we crawl publicly available movie-related information from the internet, including the director's name, cast members, and movie synopses, and use these along with the titles as seeds for DeepSeek-R1 to generate multi-speaker dialogue text.
In the design of the LLM prompt, we request that the generated dialogue text maintain logical coherence while mentioning as many names associated with the movie as possible. 
Additionally, named entities in the dialogue are also summarized by DeepSeek-R1. 
The detailed LLM prompt can be found in Appendix C2.

\begin{figure}[htp]
    \centering
    \includegraphics[width=0.47\textwidth]{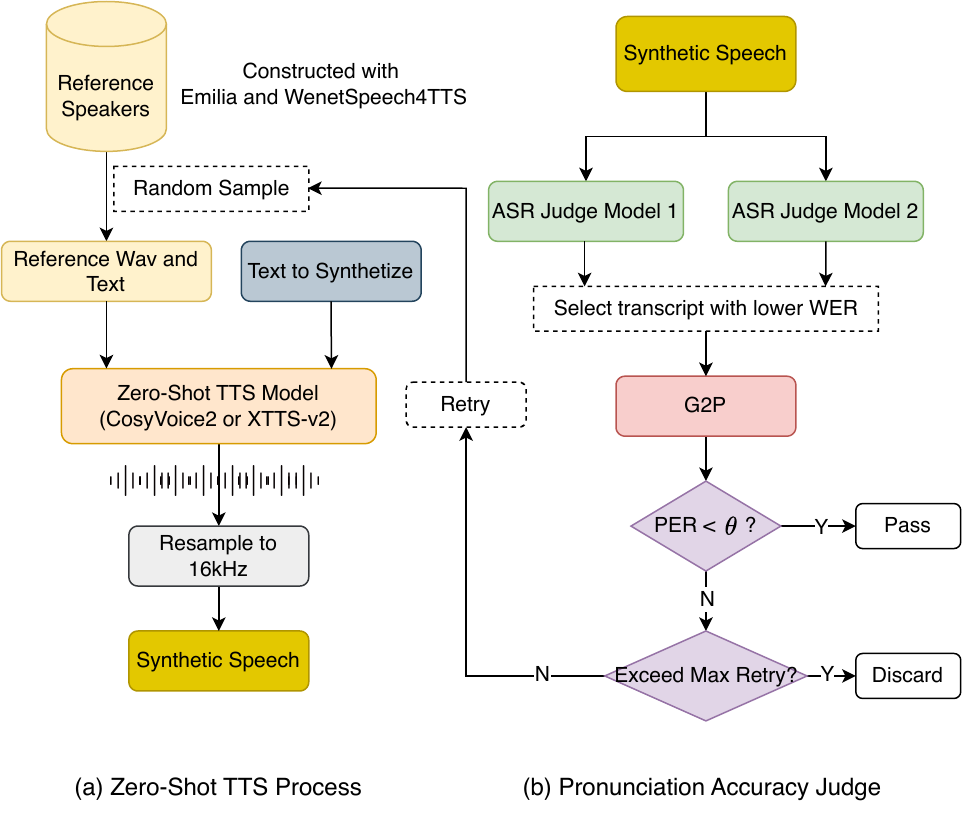}
    \caption{Overview of the Zero-Shot TTS pipeline. It includes (a) the Zero-Shot TTS process, capable of generating speaker timbre-rich and naturally fluent speeches from target texts, and (b) the pronunciation accuracy judge, which ensures the generated speech strictly follows the pronunciation of the target text, thereby ensuring the quality and reliability of ContextASR-Bench.}
    \label{fig:tts_pipeline}
\end{figure}
\subsection{Zero-Shot Speech Synthesis Pipeline}
Our principles for constructing the speech synthesis pipeline serving ASR benchmarks focus on two critical aspects: 1) Ensuring speaker timbre diversity of synthetic speech to evaluate ASR systems' robustness on speaker diversity, and 2) Guaranteeing pronunciation accuracy of synthetic speech as a fundamental ASR benchmark requirement. 
Therefore, we design a Zero-shot TTS pipeline as shown in Figure~\ref{fig:tts_pipeline}. 
Next, we will expand on it in detail based on the two aforementioned principles.

To achieve speaker diversity of synthetic speeches, we employ Zero-Shot TTS systems that allow flexible speaker timbre control by reference speech and corresponding spoken text content. 
We first construct a reference speaker database based on the WenetSpeech4TTS~\citep{wenetspeech4tts} \textit{Premium} subset, a high-quality Mandarin TTS dataset of about 945 hours, and the Emilia~\citep{emilia} dataset, a massive bilingual TTS dataset with over 100 thousand hours of speech and text data pairs. 
Specifically, both datasets are curated through DNSMOS~\citep{DNSMOS, DNSMOS_p835} score filtering (samples $\geq$ 3.0), followed by duration-based screening (3 - 20 seconds), and we randomly sampled 10,000 Mandarin and 10,000 English speech samples with corresponding transcripts as speaker timbre information. 
For speech synthesis implementation, we utilize two open-source Zero-Shot models, CosyVoice2~\citep{cosyvoice2} and XTTS-v2~\citep{xtts}, since they perform well in terms of speaker similarity and the naturalness of synthetic speech.
While prioritizing CosyVoice2 for both Mandarin and English, XTTS-v2 is only used when English speech generated by CosyVoice2 fails for the next pronunciation judge. 
This compensates for CosyVoice2's relatively weaker English synthesis capability, effectively balancing retention rates on synthetic speech data of both languages. 
At the end of the Zero-Shot TTS process, the generated speech is resampled to 16 kHz, aligning with the current standards of ASR benchmarks.

The pronunciation accuracy judge pipeline comprises two stages: First, the synthetic speech will be transcribed by the ASR systems to obtain the transcript.
Second, we employ an open-source Grapheme-to-Phoneme (G2P) converter~\footnote{https://github.com/pengzhendong/g2p-mix} to transform both transcript and target text into phoneme sequences for the PER calculation, with the threshold \(\theta\) set to 0.03. 
Specifically, to mitigate the potential bias brought by a single ASR system, we employ two ASR models for cross-verification: Sensevoice-Small~\citep{SenseVoice} for both languages, supplemented by Paraformer-Large~\citep{paraformer} for Mandarin and Whisper-Large-turbo~\citep{whisper} for English. 
For each synthetic speech, both ASR systems transcribe and obtain the transcripts.
The one with the lower WER will be chosen as the final transcription result for the following process. 
While WER remains widely adopted for TTS stability assessment, we observe its susceptibility to ASR model limitations in recognizing unusual text content (e.g, terms or named entities).
Therefore, we adopt a PER-based evaluation method to reduce the misjudgment of pronunciation accuracy caused by homophone recognition errors caused by ASR systems.
Samples failing the pronunciation accuracy judge will trigger a retry mechanism with fresh speaker sampling and resynthesis, allowing up to three retries for each entry.

For ContextASR-Speech, given the colloquial texts generated by DeepSeek-R1, all speech data are synthesized through the aforementioned pipeline. 
While the dialogue speech of ContextASR-Dialogue undergoes specialized processing, each dialogue participant is first assigned a random speaker timbre from the reference speaker database. 
Every utterance within the dialogue is individually synthesized through the Zero-Shot TTS pipeline. 
If any utterance fails in the pronunciation accuracy judgment and exceeds the maximum retry chances, the entire dialogue will be discarded. 
Conversely, all valid synthetic speech segments will be concatenated according to the sequence of the dialogue text to produce the final long audio of the multi-speaker dialogue.

\subsection{Context Evaluation and Metrics}
ContextASR-Bench aims to evaluate how world knowledge in LLMs enhances speech recognition, addressing the limitations of conventional ASR benchmarks~\citep{commonvoice, librispeech, wenetspeech, aishell1} that rigidly follow a fixed ``speech-to-text'' paradigm, lacking contextual information such as situational domains or discourse environments, thereby failing to leverage LLMs' superior contextual modeling strengths and effectively retrieve domain-specific knowledge of LLMs. 
We design the context evaluation framework, containing three evaluation settings: \textbf{Contextless}, \textbf{Coarse-grained Context}, and \textbf{Fine-grained Context}, as shown in the central part of Figure~\ref{fig:main_figure}, and use two additional metrics which are strongly related to the recognition accuracy of named entities: \textbf{NE-WER} and \textbf{NE-FNR}, as shown in the right part of Figure~\ref{fig:main_figure}.

\textbf{Context Evaluation Settings}. The \textbf{Contextless} setting closely resembles the current ASR benchmark ``speech-to-text'' paradigm, transcribing speech without any additional contextual information. 
This setting serves as a baseline applicable to both conventional ASR systems and LALMs. 
The \textbf{Coarse-grained Context} setting incorporates domain-level contextual cues into user prompts when LALMs perform speech recognition.
For ContextASR-Speech set, this involves providing domain labels for each data entry, while for ContextASR-Dialogue set, it refers to the movie title relevant to the dialogue.
This setting evaluates LALMs' capability to retrieve domain-specific knowledge from their internal world knowledge when given vague contextual hints, thereby enhancing speech understanding. 
We posit that LALMs' true value in speech recognition lies in their ability to generalize across domains through coarse-grained prompting, which is also the Coarse-grained Context setting designed to assess.
The \textbf{Fine-grained Context} setting employs precise prior knowledge injection by incorporating terms or named entities within the speech text content into the user prompt.
This setting simulates practical scenarios requiring user-customized recognition capabilities, particularly for recognizing organization-specific jargon or personal idiosyncratic expressions. 

\textbf{Evaluation Metrics}.
Existing ASR benchmarks rely on WER, calculated as
\(\frac{S + I + D}{T}\),
where \(S\), \(I\), and \(D\) represent substitution, insertion, and deletion errors when calculating edit distance between ground-truth text and transcript, and \(T\) is the total word count of ground-truth text. 
However, WER treats all words equally, conflicting with human evaluation priorities that emphasize critical content, such as named entities, technical terms, over functional words, such as tone words or pronouns.
To bridge this gap, we introduce two entity-centric metrics, NE-WER and NE-FNR. 
The \textbf{NE-WER}, similar to the biased WER (B-WER) which focuses on the biased keywords~\citep{lcb_net,context_with_inter_bias}, follows the same calculation formula as WER, but exclusively on entity spans using fuzzy matching with an edit distance tolerance of at most \(\lceil \frac{n}{2} \rceil - 1\), where n is the number of words in the entity. For example, a 5-word entity allows up to \(\lceil \frac{5}{2} \rceil - 1\), that is 2 errors.
It effectively focuses on the evaluation of entity recognition accuracy. 
Additionally, the more stringent \textbf{NE-FNR} adopts exact matching to quantify entity miss rates, calculated as 
\(1 - \frac{H}{N}\),
where \(H\) and \(N\) denote recognized and ground-truth entity counts. 
NE-FNR inversely corresponds to the Recall commonly used in classification tasks, providing a stringent measure of entity detection precision. 
Together, NE-WER and NE-FNR offer complementary insights: NE-WER evaluates error patterns in entity recognition, while NE-FNR assesses recognition reliability of the whole entity, critical for applications requiring high-precision entity transcribing.

\section{Experiments and Analyses}
\label{sec:experiments}
\newcolumntype{Z}{>{\centering\arraybackslash}X}
To highlight our proposed ContextASR-Bench in assessing how LLMs' world knowledge and context modeling capabilities enhance contextual speech recognition, we conduct comprehensive evaluations. 
We evaluate conventional ASR models, including Paraformer-Large~\citep{paraformer}, SenseVoice-Small~\citep{SenseVoice}, Whisper-Large-v3 and turbo~\citep{whisper}, FireredASR-AED-L and FireredASR-LLM-L~\citep{FireRedASR}, Dolphin-Base and Small~\citep{dolphin}, as well as LALMs, including Qwen2-Audio~\citep{qwen2audio}, Qwen2.5-Omni~\citep{qwen25omni}, Baichuan-Audio~\citep{baichuan-audio}, Baichuan-Omni-1.5~\citep{baichuan-omni-1d5}, and Kimi-Audio~\citep{kimi-audio}.
All user prompts for LALMs under three context evaluation settings can be found in Appendix D.

\begin{figure}[tp]
    \centering
    \includegraphics[width=0.47\textwidth]{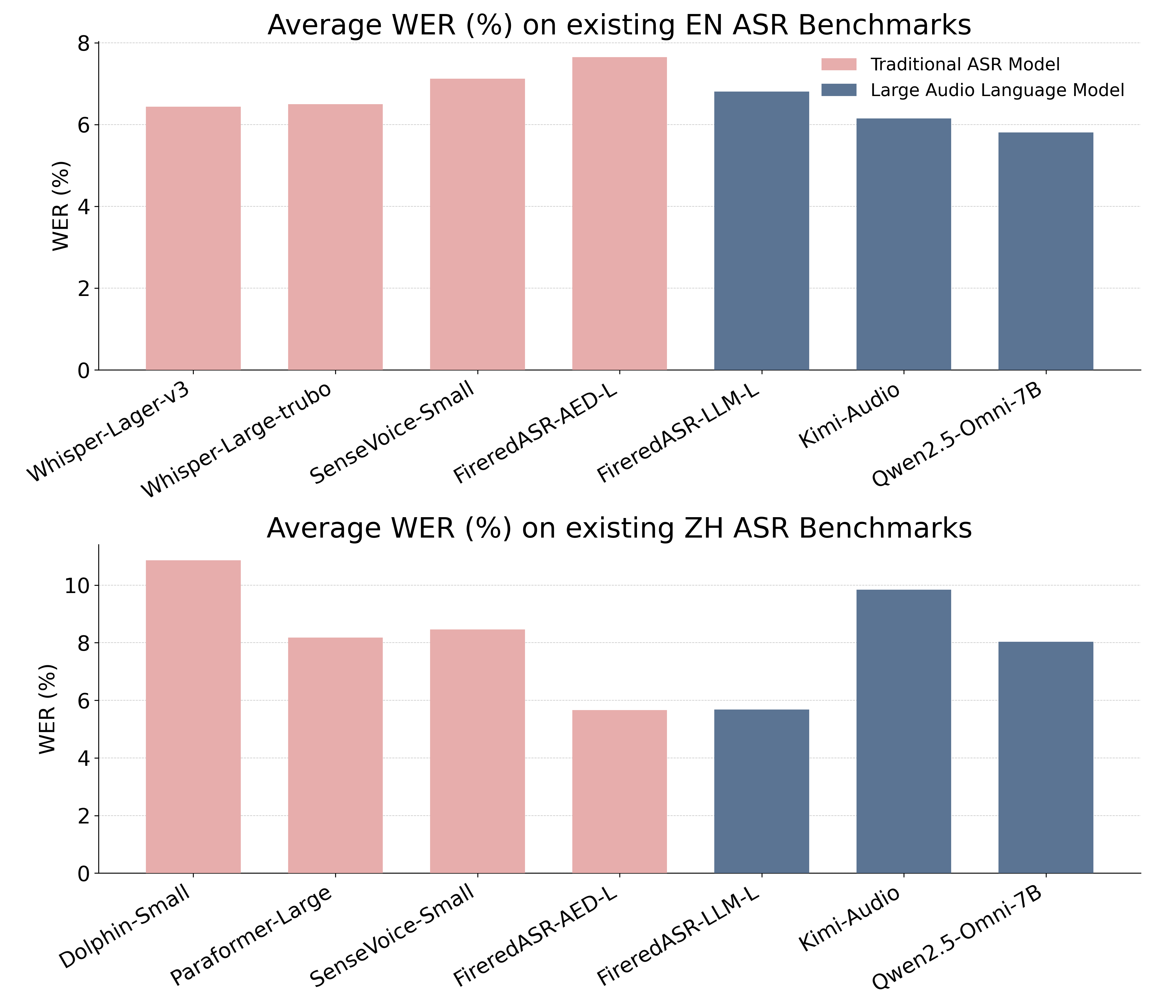}
    \caption{Comparison of average WER between conventional ASR models and LALMs on existing open-source Mandarin~(ZH) and English~(EN) ASR benchmarks.}
    \label{fig:existing_bench}
\end{figure}

\begin{table*}[ht]
  \centering
  \fontsize{9}{11}\selectfont{
  \begin{tabularx}{\linewidth}{@{} 
      c   
      r   
      c   
      Z   
      Z   
      Z
      Z
    @{}}
    \toprule
    \multirow{4}{*}{\textbf{Model}} 
      & \multirow{4}{*}{\textbf{Size}} 
      & \multirow{4}{*}{\textbf{Context}} 
      & \textbf{ContextASR-} &\textbf{ContextASR-} & \textbf{ContextASR-} &\textbf{ContextASR-} \\
      & & & \textbf{Speech-EN} &\textbf{Dialogue-EN} & \textbf{Speech-ZH} &\textbf{Dialogue-ZH} \\
    \cmidrule(lr){4-7}
      & & 
      & \textbf{ WER $ \mid $ NE-WER }
      & \textbf{ WER $ \mid $ NE-WER $ \mid $}
      & \textbf{ WER $ \mid $ NE-WER $ \mid $}
      & \textbf{ WER $ \mid $ NE-WER $ \mid $}\\
      & & 
      & \textbf{NE-FNR (\%) $\downarrow$}
      & \textbf{NE-FNR (\%) $\downarrow$}
      & \textbf{NE-FNR (\%) $\downarrow$}
      & \textbf{NE-FNR (\%) $\downarrow$}\\

    \midrule
    \multicolumn{7}{c}{\textbf{Automatic Speech Recognition Models (ASRs)}}  \\
    \midrule
    Paraformer-Large & 220M & \multirow{8}{*}{/} 
      & 34.33 $ \mid $ 76.71 $ \mid $ 91.44 
      & 27.79 $ \mid $ 78.22 $ \mid $ 82.81 
      & ~5.62 $ \mid $ 28.68 $ \mid $ 55.71
      & ~5.97 $ \mid $ 36.62 $ \mid $ 52.45\\
    Sensevoice-Small & 234M &    
      & 15.72 $ \mid $ 56.78 $ \mid $ 77.96 
      & 11.45 $ \mid $ 55.67 $ \mid $ 61.16
      & ~6.02 $ \mid $ 32.79 $ \mid $ 65.18	
      & ~6.35 $ \mid $ 39.67 $ \mid $ 58.41\\
    Whisper-Large-v3 & 1.5B &     
      & ~9.36 $ \mid $ 29.56 $ \mid $ 39.89 
      & ~9.62 $ \mid $ 33.55 $ \mid $ 35.24
      & 13.62 $ \mid $ 46.58 $ \mid $ 77.35	
      & ~9.05 $ \mid $ 44.79 $ \mid $ 62.33 \\
    Whisper-Large-turbo & 809M &   
      & ~9.84 $ \mid $ 32.10 $ \mid $ 44.01 
      & ~9.36 $ \mid $ 34.68 $ \mid $ 36.66 
      & 14.70 $ \mid $ 49.47 $ \mid $ 82.24
      & 10.10 $ \mid $ 47.16 $ \mid $ 66.58\\
    Dolphin-Base & 140 M & 
    &  - $ \mid $ - $ \mid $ -
    &  - $ \mid $ - $ \mid $ -
    & 12.95 $ \mid $ 50.42 $ \mid $ 85.79
    & 10.18 $ \mid $ 45.88 $ \mid $ 64.13 \\
    Dolphin-Small & 372 M & 
    &  - $ \mid $ - $ \mid $ -
    &  - $ \mid $ - $ \mid $ -
    & 10.68 $ \mid $ 46.29 $ \mid $ 82.49
    & ~7.73 $ \mid $ 41.48 $ \mid $ 58.37 \\
    FireredASR-AED-L   & 1.1B &     
      & 13.72 $ \mid $ 48.88 $ \mid $ 69.14 
      & 15.28 $ \mid $ 51.88 $ \mid $ 57.03
      & ~4.00 $ \mid $ 22.81 $ \mid $ 41.33
      & ~4.43 $ \mid $ 31.19 $ \mid $ 41.30\\
    FireredASR-LLM-L   & 8.3B &  
      & ~6.93 $ \mid $ 23.69 $ \mid $ 32.74 
      & ~6.50 $ \mid $ 30.59 $ \mid $ 32.18 
      & ~2.83 $ \mid $ 16.14 $ \mid $ 26.75 
      & ~\textbf{3.24} $ \mid $ 23.28 $ \mid $ 30.08\\
    \midrule
    \multicolumn{7}{c}{\textbf{Large Audio Language Models (LALMs)}} \\
    \midrule
    \multirow{3}{*}{Qwen2-Audio} & \multirow{3}{*}{8.4B} & /   
      & 13.56 $ \mid $ 38.95 $ \mid $ 52.29 
      & 14.16 $ \mid $ 42.25 $ \mid $ 44.92 
      & 10.14 $ \mid $ 28.73 $ \mid $ 41.45
      & ~7.34 $ \mid $ 27.85 $ \mid $ 35.08 \\
      &  & Coarse   
      & 13.41 $ \mid $ 38.34 $ \mid $ 51.55
      & 13.85 $ \mid $ 37.88 $ \mid $ 40.01
      & 10.17 $ \mid $ 28.72 $ \mid $ 41.42
      & ~7.67 $ \mid $ 27.61 $ \mid $ 34.61 \\
      &  & Fine 
      & 11.49 $ \mid $ 27.27 $ \mid $ 35.08 
      & 13.99 $ \mid $ 33.02 $ \mid $ 32.92 
      & ~9.92 $ \mid $ 24.10 $ \mid $ 30.02
      & ~7.00 $ \mid $ 22.76 $ \mid $ 26.17\\
    \cline{4-7}
    \multirow{3}{*}{Baichuan-Audio} & \multirow{3}{*}{10.4B} & / 
      & 13.02 $ \mid $ 20.64 $ \mid $ 26.84 
      & ~9.46 $ \mid $ 23.27 $ \mid $ 23.26 
      & ~7.30 $ \mid $ \textbf{14.19} $ \mid $ \textbf{17.64}
      & ~5.83 $ \mid $ 29.14 $ \mid $ 34.71\\
      &  & Coarse   
      & ~9.33 $ \mid $ 19.44 $ \mid $ 25.84
      & ~6.46 $ \mid $ 18.62 $ \mid $ 17.78
      & ~3.07 $ \mid $ \textbf{12.73} $ \mid $ \textbf{17.12}
      & ~3.82 $ \mid $ 25.29 $ \mid $ 29.61 \\
      &  & Fine$^\ast$ 
      & 7.52 $ \mid $ ~\textbf{5.87} $ \mid $ \textbf{4.55}
      & ~5.66 $ \mid $ \textbf{10.01} $ \mid $ \textbf{3.64} 
      & ~2.16 $ \mid $ ~\textbf{6.65} $ \mid $ ~\textbf{2.35}
      & ~2.96 $ \mid $ \textbf{11.48} $ \mid $ ~\textbf{3.94}\\
    \cline{4-7}
    \multirow{3}{*}{Kimi-Audio} & \multirow{3}{*}{9.8B} & /   
      & ~\textbf{4.09} $ \mid $ \textbf{14.33} $ \mid $ \textbf{19.53} 
      & ~\textbf{4.58} $ \mid $ \textbf{18.19} $ \mid $ \textbf{17.74}
      & ~2.60 $ \mid $ 16.49 $ \mid $ 27.84
      & ~3.44 $ \mid $ \textbf{22.33} $ \mid $ \textbf{27.68} \\
      &  & Coarse   
      & ~\textbf{4.47} $ \mid $ \textbf{13.88} $ \mid $ \textbf{18.60}
      & ~\textbf{4.78} $ \mid $ \textbf{17.28} $ \mid $ \textbf{16.54}
      & ~\textbf{2.47} $ \mid $ 15.75 $ \mid $ 26.12
      & ~3.34 $ \mid $ \textbf{21.31} $ \mid $ \textbf{25.94} \\
      &  & Fine 
      & \textbf{2.90} $ \mid $ ~6.68 $ \mid $ ~8.01 
      & ~\textbf{4.67} $ \mid $ 13.50 $ \mid $ 11.31
      & ~1.95 $ \mid $ 11.13 $ \mid $ 15.28
      & 2.90 $ \mid $ 15.91 $ \mid $ 16.68\\
    \cline{4-7}
    \multirow{3}{*}{Baichuan-Omni-1.5} & \multirow{3}{*}{11B} & /   
      & 10.65 $ \mid $ 23.17 $ \mid $ 30.15 
      & 11.05 $ \mid $ 29.78 $ \mid $ 30.81 
      & ~3.42 $ \mid $ 14.88 $ \mid $ 21.18	
      & ~5.42 $ \mid $ 33.44 $ \mid $ 41.88 \\
      &  & Coarse   
      & 11.17 $ \mid $ 23.06 $ \mid $ 29.88 
      & ~9.86 $ \mid $ 26.11 $ \mid $ 25.97 
      & ~3.73 $ \mid $ 14.90 $ \mid $ 20.88
      & ~5.12 $ \mid $ 30.44 $ \mid $ 37.19 \\
      &  & Fine$^\ast$ 
      & 8.16 $ \mid $ ~7.69 $ \mid $ ~6.53 
      & ~9.91 $ \mid $ 14.40 $ \mid $ ~5.54 
      & ~2.98 $ \mid $ ~8.39 $ \mid $ ~4.71
      & ~5.00 $ \mid $ 16.83 $ \mid $ 7.84\\
    \cline{4-7}
    \multirow{3}{*}{Qwen2.5-Omni-3B} & \multirow{3}{*}{5.4B} & /   
      & ~6.19 $ \mid $ 20.52 $ \mid $ 28.26 
      & ~5.94 $ \mid $ 28.29 $ \mid $ 29.28 
      & ~3.48 $ \mid $ 20.68 $ \mid $ 37.44	
      & ~4.35 $ \mid $ 30.07 $ \mid $ 40.51 \\
      &  & Coarse   
      & ~6.30 $ \mid $ 20.62 $ \mid $ 28.33 
      & ~5.73 $ \mid $ 26.65 $ \mid $ 27.28 
      & ~3.34 $ \mid $ 19.82 $ \mid $ 35.39
      & ~4.05 $ \mid $ 27.50 $ \mid $ 36.03 \\
      &  & Fine 
      & 3.99 $ \mid $ ~7.80 $ \mid $ ~9.69 
      & ~4.83 $ \mid $ 14.36 $ \mid $ 12.85 
      & ~2.13 $ \mid $ 10.55 $ \mid $ 14.11
      & ~3.12 $ \mid $ 15.07 $ \mid $ 15.17\\
    \cline{4-7}
    \multirow{3}{*}{Qwen2.5-Omni-7B} & \multirow{3}{*}{10.1B} & /   
      & ~5.60 $\mid$ 16.07 $ \mid $ 21.33
      & ~5.78 $ \mid $ 20.60 $ \mid $ 20.50 
      & ~\textbf{2.59} $ \mid $ 19.05 $ \mid $ 33.88
      & ~3.70 $ \mid $ 26.52 $ \mid $ 34.52 \\
      &  & Coarse   
      & ~5.56 $ \mid $ 15.93 $ \mid $ 21.13 
      & ~6.21 $ \mid $ 18.88 $ \mid $ 18.42 
      & ~3.14 $ \mid $ 18.26 $ \mid $ 31.99 
      & ~\textbf{3.28} $ \mid $ 23.76 $ \mid $ 29.77 \\
      &  & Fine
      & 3.96 $ \mid $ ~7.38 $ \mid $ ~8.72 
      & ~5.32 $ \mid $ 11.83 $ \mid $ ~9.24
      & ~\textbf{1.84} $ \mid $ ~9.80 $ \mid $ 12.19
      & ~\textbf{2.40} $ \mid $ 14.06 $ \mid $ 13.17 \\
    \bottomrule
    \end{tabularx}
  }
  \caption{
      Results of all evaluated models on ContextASR-Bench. 
      All models are classified into ASR models and LALMs, based on whether they have instruction following capacity and can be evaluated under all context evaluation settings.
      ``Size'' refers to the total number of parameters in the model. ``Context'' refers to the context evaluation setting on which the model is evaluated, where ``/'', ``Coarse'', and ``Fine'' indicate the Contextless setting, Coarse-grained Context setting, and Fine-grained Context setting. 
      Severe hallucination is observed in the transcription results under the Fine-grained Context setting marked with an ``$\ast$''.
    }
    \label{tab:main_results}
\end{table*}

\subsection{Results on Existing ASR Benchmarks}
To thoroughly demonstrate that existing ASR benchmarks fail to unveil the improvements brought by LLM to speech recognition tasks, we select several representative conventional ASR systems and LLM-based ASR systems and test them on 21 English and 54 Mandarin open-source benchmarks. 
The average WER of each model across all datasets is shown in Figure~\ref{fig:existing_bench}, and detailed results can be found in Appendix B.
The WERs for ASR systems with or without LLM on these open-source benchmarks show little difference. 
Even the Paraformer-Large model, with only 220M parameters, outperformed Kimi-Audio-7B on Mandarin benchmarks, which defies intuition.
It can be attributed to two main reasons:
1) The text domain in open-source benchmarks is narrow, with frequent casual conversational context, which limits the applicability of LLMs' extensive domain knowledge and context-understanding capabilities.
2) The WER calculations assign equal weight to all tokens, which fails to effectively highlight the areas where LLMs advantage, such as named entities or terms.
These results strongly support the necessity of ContextASR-Bench.

\subsection{Evaluation Settings on ContextASR-Bench}
Considering the recordings in ContextASR-Bench are longer than the 30s input limit of many conventional ASR models, such as the Whisper and FireredASR series, we segment each utterance with an open-source voice activity detection (VAD) tool~\citep{funasr}, merge the short resulting chunks and ensure no segment exceeds 30 seconds, transcribe each segment separately, and then concatenate the partial transcripts. For ContextASR-Dialogue, with a 150s average speech duration, current open-source LALMs often exhibit severe hallucinations or truncations, so we also apply the VAD preprocessing to obtain reliable evaluation results.

\subsection{Results on ContextASR-Bench}
\subsubsection{Conventional ASR Models vs. LALMs}
Table~\ref{tab:main_results} presents all the test results of evaluated ASR systems on ContextASR-Bench. 
It is evident that ASR systems without LLMs generally have NE-FNR rates exceeding 50\% on both ContextASR-Speech set and ContextASR-Dialogue set.
Even FireredASR-AED-L, the current SOTA ASR model for Mandarin, shows an NE-FNR exceeding 40\% on ContextASR-Bench. 
In contrast, the LALM models perform evidently better even in the Contextless setting compared to conventional ASR models. 
Qwen2.5-Omni-7B exhibits a relative reduction of 39.9\% in WER and 42\% in NE-FNR on ContextASR-Dialogue (EN) compared to the Whisper-Large-V3, the current SOTA English ASR model. 
However, these two models only show a 9.8\% difference on existing English ASR benchmarks. 
The above indicates: 1) ContextASR-Bench has a greater distinction capability between conventional ASR models and LALMs compared to existing ASR benchmarks, highlighting that the strong world knowledge and context learning capabilities of LALMs are important for contextual speech recognition. 
2) LALM models can still perform generally well in the Contextless setting, leveraging the massive text training data and world knowledge built on it.

\subsubsection{Coarse- and Fine-grained Context}
\begin{figure}[tp]
    \centering
    \includegraphics[width=0.475\textwidth]{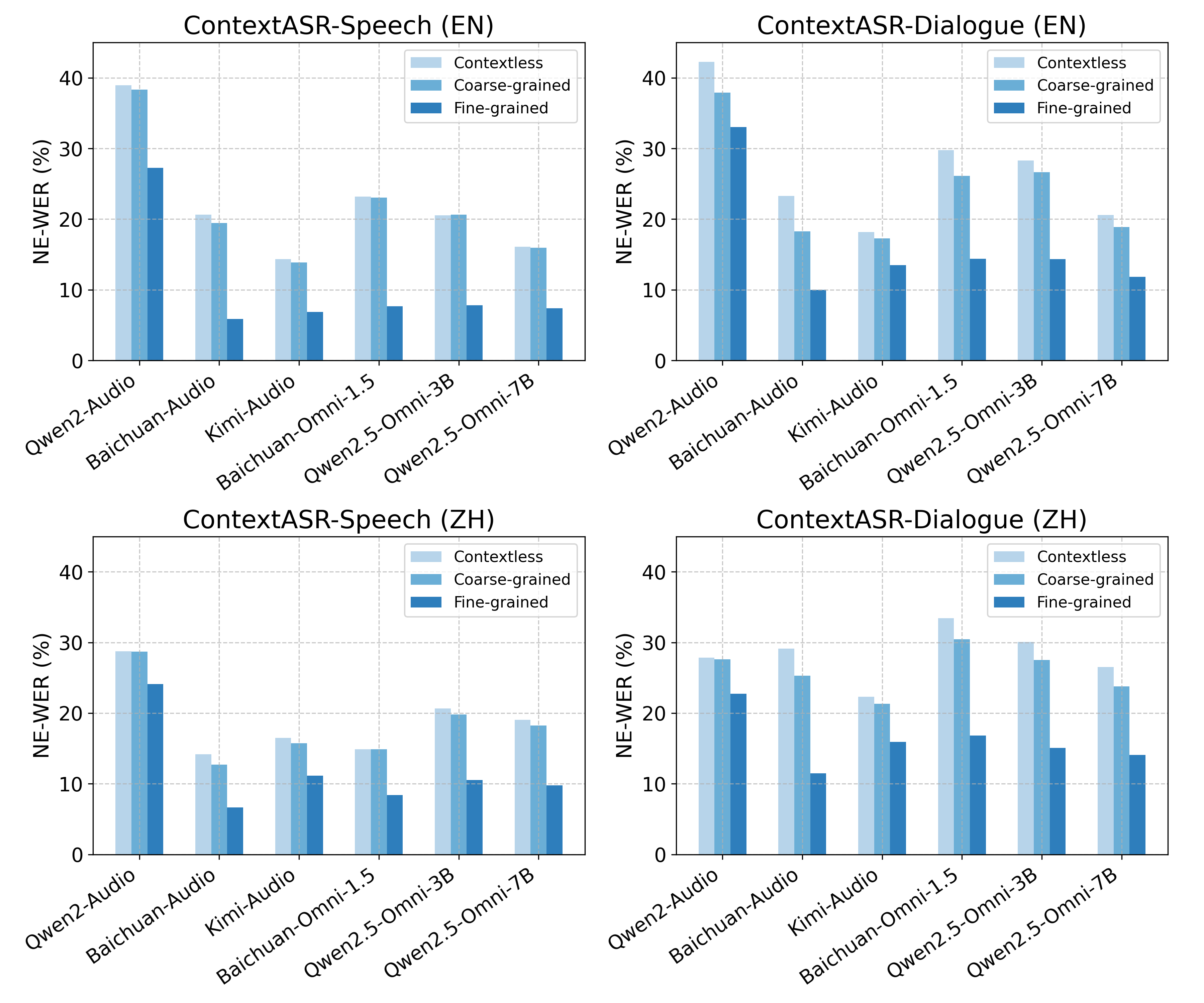}
    \caption{The NE-WER (\%) of LALMs on ContextASR-Bench under Contextless, Coarse-grained Context and Fine-grained Context evaluation settings.}
    \label{fig:context_promote_on_newer}
\end{figure}

Figure~\ref{fig:context_promote_on_newer} compares NE-WER metrics of LALMs evaluated under Coarse-grained and Fine-grained Context settings with the Contextless setting. 
We can notice that LALMs show more obvious reductions in NE-WER on ContextASR-Dialogue set compared to ContextASR-Speech set under the Coarse-grained Context setting.
ContextASR-Speech set uses the domain label of speech text content as coarse-grained context, which differs in precision from the movie title used in ContextASR-Dialogue set; 
Domain labels are more generalized, whereas movie titles are more specific. 
This indicates that current LALMs still have limited capability in retrieving specific knowledge to enhance the speech recognition task through broad contexts, such as domain labels. 
While under the Fine-grained Context setting, LALMs show a significant reduction in NE-WER, lining up with expectations. 
However, we observe that under this setting, some LALMs begin to generate severe hallucinations, manifested as repeating only emitting entities within the text prompt when transcribing speech. 
As a result, although the NE-WER and NE-FNR metrics show big decreases, the WER did not exhibit a similar reduction. 
For instance, Baichuan-Audio and Baichuan-Omni-1.5, these two models achieve much lower NE-FNR than other LALMs on both ContextASR-Speech and ContextASR-Dialogue sets under the Fine-grained Context setting.
However, their WERs are noticeably higher than others. 
The appearance of hallucinations under the Fine-grained Context setting indicates that the model is paying too much attention to the prompt while somehow ignoring the auditory modality input. 
It suggests that in the contextual speech recognition task, balancing the model's attention to text modality context information and audio modality is crucial for achieving stable and reliable transcriptions.

\section{Related Work}

Recently, ASR research has been driven by a diverse array of open-source corpora spanning multiple domains or languages. 
\textbf{General domain} benchmarks include THCHS‐30~\citep{THCHS-30}, focusing on Mandarin read speech; LibriSpeech~\citep{librispeech}, the standard English audiobook corpus; AISHELL‐1~\citep{aishell1} and AISHELL‐2~\citep{Aishell2} for indoor and mobile‐device recordings; SPGISpeech~\citep{SPGISpeech} for financial telephony; Common Voice~\citep{commonvoice} for crowdsourced accent variation; and large‐scale web‐sourced collections such as WenetSpeech~\citep{wenetspeech}, GigaSpeech~\citep{gigaspeech}, and the unified SpeechIO leaderboard\footnote{\url{https://github.com/SpeechColab/Leaderboard}} with massive test subsets.
\textbf{Multilingual} evaluations are supported by FLEURS~\citep{FLEURS}, Multilingual LibriSpeech\citep{MLS}.
SEAME\citep{SEAME} and the ASRU~\citep{ASRU2019} benchmarks are designed for \textbf{code-switching} ASR. 
\textbf{Accent and dialect} diversity are examined in KeSpeech~\citep{kespeech}, covering eight regional Mandarin variants, and VoxPopuli~\citep{voxpopuli}, focusing on accented news broadcasts. 
\textbf{Far‐field and multi‐speaker} scenarios are addressed by AISHELL‐4~\citep{aishell4}, AliMeeting~\citep{alimeeting}, and AISHELL-5~\citep{aishell-5}, providing multi‐channel meeting recordings. 
\textbf{Scenario‐specific} benchmarks such as SlideSpeech~\citep{Slidespeech} and TED‐LIUM~\citep{TED-lium} cater to slide‐synchronized presentations and TED talks.

\section{Conclusion}
In this paper, we introduce ContextASR-Bench, the first massive contextual speech recognition benchmark specifically designed for evaluating linguistic knowledge and contextual capture capabilities of Automatic Speech Recognition (ASR) models.
This benchmark encompasses over 40,000 data entries across more than 10 domains, and each item provides additional context, such as the domain it belongs to and the named entities it contains, enabling a thorough evaluation of how effectively models can exploit contextual cues to improve ASR accuracy.
Extensive experiments reveal that Large Audio Language Models (LALMs) outperform conventional ASR models considerably, due to their inherent extensive world knowledge and context-learning abilities. 
Nevertheless, current LALMs still struggle in contextual ASR, indicating ample room for improvement.
We open-source ContextASR-Bench in the hope of stimulating further research on how LALMs can better leverage additional context information and accelerating progress toward truly universal ASR systems across all domains.

\bibliography{aaai2026}


\appendix
\section{A. Details for NER Datasets}
\label{appendix:ner_dataset}
In Section 2.1, we described the process of generating text corpora for ContextASR-Bench, where we obtain domain-specific named entities from the collected open-source NER datasets as seeds for DeepSeek-R1 to generate entity-rich colloquial text of ContextASR-Speech.
Table~\ref{tab:ner_datasets} summarizes the statistics of the NER datasets we used, including CMeEE~\citep{ner_cmeee}, IMCS21 Task 1~\citep{ner_imcs21}, CLUENER~\citep{ner_cluener}, MSRA~\citep{ner_msra}, NLPCC2018 Task 4~\citep{ner_nlpcc}, CCFBDCI~\footnote{https://www.datafountain.cn/competitions/510}, MMC~\citep{ner_mmc}, E-Commerce~\citep{ner_ECommerce}, Resume~\citep{ner_resume}, 
Bank~\footnote{https://www.heywhale.com/mw/dataset/\\617969ec768f3b0017862990}, 
FNED~\citep{ner_FNED}, DLNER~\citep{ner_dlner} datasets.

\begin{table*}[htp]
  \centering
  \fontsize{9}{11}{
  \begin{tabularx}{\linewidth}{llcc}
    \toprule
    \textbf{Dataset} & \textbf{Description} & \textbf{Size} & \textbf{Entity Category} \\
    \midrule
    CMeEE
      & CBLUE Chinese Medical Entity Recognition
      & 20,000
      & 9 \\
    IMCS21 Task 1
      & Dataset from the 1st CCL2021 Intelligent Dialogue Diagnosis Challenge
      & 98,452
      & 5 \\
    CLUENER
      & A fine-grained dataset from Sina News RSS
      & 12,091
      & 10 \\
    MSRA
      & Microsoft Research Asia open-source NER
      & 48,442
      & 3 \\
    NLPCC2018 Task 4
      & Task-oriented Dialogue System NER
      & 21,352
      & 15 \\
    CCFBDCI
      & Chinese NER Algorithm Robustness Evaluation
      & 15,723
      & 4 \\
    MMC
      & MMC AI-assisted Knowledge Graph Construction Challenge
      & 3,498
      & 18 \\
    E-Commerce
      & E-commerce-oriented NER
      & 7,998
      & 4 \\
    Resume
      & Executives’ Resumes from Chinese Listed Companies
      & 4,761
      & 8 \\
    Bank
      & Banking Loan Data NER
      & 10,000
      & 4 \\
    FNED
      & Domain Event Detection under High Robustness Requirements
      & 10,500
      & 7 \\
    DLNER
      & Discourse-level NER
      & 28,897
      & 9 \\
    \bottomrule
  \end{tabularx}
  }
  \caption{Details of open-source NER datasets used as seeds for ContextASR-Bench corpora generation prompt.}
  \label{tab:ner_datasets}
\end{table*}

\section{B. Detailed Results on Existing Open-Source ASR Benchmarks}
\label{appendix:open-source_asr_benchmarks}
While LALMs that incorporate LLMs deliver qualitative gains in general audio understanding and reasoning compared with small-parameter models, they have not yet demonstrated the same striking advantage in ASR as they have in other tasks. 
We contend that this discrepancy does not mean LLMs are less beneficial to ASR; rather, current ASR benchmarks fail to reveal the improvements LLMs can bring.
To thoroughly demonstrate this, we select several representative conventional ASR systems and LLM-based ASR systems and test them on 21 English and 54 Mandarin existing open-source benchmarks.
Table~\ref{tab:asr_en_wer} and Table~\ref{tab:asr_zh_wer} show the detailed results on English and Mandarin benchmarks, respectively.
The experiments reveal that on existing open-source English ASR benchmarks, the LALM with the lowest average WER, Qwen2.5-Omni, offers less than a 10\% relative improvement over Whisper-Large-v3, despite having more than five times as many parameters. 
On the Mandarin benchmarks, the top-performing LALM, FireredASR-LLM-L, even underperforms compared with FireredASR-AED-L.
\begin{table*}[htp]
  \centering
  \fontsize{9}{11}\selectfont{
  \begin{tabularx}{\linewidth}{@{}l *{7}{c}@{}}
    \toprule
    Dataset & SV-Small & FASR-AED-L
            & W-L-turbo & W-L-v3 
            & FASR-LLM-L & Qwen2.5-Omni & Kimi-Audio \\
    \midrule
    CV\_V17.0\_EN\_DEV      & 13.13 & 13.88 &  9.05 &  8.15 & 10.94 &  5.93 &  6.13 \\
    CV\_V17.0\_EN\_TEST     & 14.98 & 17.55 & 12.04 & 10.56 & 15.00 &  7.72 &  8.35 \\
    FLEURS\_EN-US\_DEV                &  8.70 &  8.09 &  4.97 &  4.63 &  4.88 &  4.20 &  5.20 \\
    FLEURS\_EN-US\_TEST               &  7.83 &  7.70 &  4.84 &  4.59 &  4.65 &  3.81 &  4.63 \\
    GS\_V1.0.0\_DEV           & 11.61 & 10.11 & 11.11 & 11.45 &  9.55 & 11.06 & 10.44 \\
    GS\_V1.0.0\_TEST          & 11.73 & 10.17 & 10.55 & 10.64 &  9.66 & 11.02 & 10.06 \\
    LS\_DEV\_CLEAN           &  3.49 &  1.82 &  2.25 &  2.24 &  1.58 &  1.55 &  1.60 \\
    LS\_DEV\_OTHER           &  6.99 &  4.22 &  4.25 &  3.96 &  3.06 &  3.23 &  2.82 \\
    LS\_TEST\_CLEAN          &  3.26 &  1.94 &  2.37 &  2.15 &  1.65 &  1.74 &  1.58 \\
    LS\_TEST\_OTHER          &  7.30 &  4.39 &  4.36 &  3.96 &  3.65 &  3.47 &  2.93 \\
    MLS\_EN\_TEST                     & 10.22 &  7.01 &  5.39 &  5.21 &  5.22 &  5.45 &  4.82 \\
    SLIDE\_SPEECH\_DEV               & 10.44 &  7.67 &  9.17 &  9.53 &  7.54 &  8.42 &  8.18 \\
    SLIDE\_SPEECH\_TEST              & 11.23 &  8.18 & 10.81 & 10.75 &  8.11 &  9.81 &  8.99 \\
    SPGISPEECH\_DEV                  &  3.49 &  5.37 &  3.33 &  3.46 &  4.70 &  2.25 &  4.00 \\
    SPGISPEECH\_TEST                 &  3.50 &  5.28 &  3.27 &  3.36 &  4.60 &  2.27 &  3.91 \\
    TEDLIUM3\_LEGACY\_DEV   &  4.53 &  4.75 &  4.40 &  4.15 &  4.11 &  3.68 &  3.44 \\
    TEDLIUM3\_LEGACY\_TEST  &  4.16 &  3.96 &  4.27 &  4.60 &  3.84 &  3.93 &  3.36 \\
    VP\_V1.0\_EN\_ACCENT\_TEST & 14.16 & 14.25 & 18.90 & 18.77 & 13.96 & 23.71 & 16.52 \\
    VP\_V1.0\_EN\_DEV          & 10.82 & 10.73 &  9.99 &  9.72 & 10.32 &  6.70 &  8.90 \\
    VP\_V1.0\_EN\_TEST         & 10.45 & 10.61 & 10.67 &  9.14 &  9.96 &  6.51 &  8.91 \\
    \midrule
    Overall                           &  7.12 &  7.65 &  6.50 &  6.44 &  6.81 &  5.81 &  6.15 \\
    \bottomrule
  \end{tabularx}
  }
  \caption{
    The WER (\%) results of conventional ASR systems and Large Audio–Language Models (LALMs) on existing open-source English ASR benchmarks.``Overall" denotes the average WER each model achieves across all benchmark test sets. ``SV-Small" refers to SenseVoice-Small, ``FASR" to FireredASR, and ``W-L-turbo" and ``W-L-v3" to Whisper-Large-turbo and Whisper-Large-v3, respectively. ``CV" refers to CommonVoice, ``GS" to GigaSpeech, ``LS" to LibriSpeech, and ``VP" to VoxPopuli.
  }
  \label{tab:asr_en_wer}
\end{table*}

\begin{table*}[htbp]
  \centering
  \fontsize{9}{11}\selectfont{
  \begin{tabularx}{\linewidth}{@{}l *{7}{c}@{}}
    \toprule
    Dataset & SV-Small & PF-Large & FASR-AED-L
            & Dolphin-Small & FASR-LLM-L & Qwen2.5-Omni & Kimi-Audio \\
    \midrule

    AS1\_TEST            & 3.01  & 1.93  & 0.55   & 3.33  & 0.73  & 1.62  & 0.76  \\
    AS2\_AOS\_TEST   & 3.94  & 3.08  & 2.76   & 4.74  & 2.50  & 2.76  & 2.63  \\
    AS2\_IOS\_TEST       & 3.81 & 2.84  & 2.52  & 4.42  & 2.16  & 2.59  & 2.84  \\
    AS2\_MIC\_TEST       & 3.88  & 3.01  & 2.81   & 4.78  & 2.49  & 2.63  & 2.76  \\
    AS4\_TEST           & 16.59  & 17.13  & 11.79   & 20.15  & 12.06  & 19.26  & 20.00  \\
    AM\_EVAL\_FAR & 24.21  & 21.84  & 14.22   & 30.68  & 14.87  & 26.81  & 26.06  \\
    AM\_EVAL\_NEAR & 5.55  & 5.15  & 3.34   & 6.28  & 3.97  & 5.56  & 6.98  \\
    AM\_TEST\_FAR & 25.42  & 23.12  & 15.44 & 32.61  & 16.35  & 29.86  & 29.30  \\
    AM\_TEST\_NEAR & 7.05  & 6.47  & 4.09  & 8.38  & 4.89  & 6.87  & 8.35  \\
    ASRU\_TEST               & 8.12  & 5.34  & 6.60  & 9.20   & 5.71  & 8.39  & 7.21  \\
    CV\_V17.0\_ZH\_DEV  & 13.47  & 12.95  & 7.15  & 17.95  & 7.20  & 8.38  & 9.45  \\
    CV\_V17.0\_ZH\_TEST & 10.57  & 10.24  & 3.39   & 11.53  & 3.51  & 5.06  & 6.06  \\
    FLEURS\_CMN\_DEV         & 3.56  & 3.34  & 3.20   & 4.07  & 2.41  & 2.31  & 2.28  \\
    FLEURS\_CMN\_TEST        & 4.16  & 3.80  & 3.64   & 4.48  & 2.54  & 2.59  & 2.52  \\
    KESPEECH\_DEV            & 8.43  & 9.53  & 3.82   & 8.61  & 3.17  & 5.58  & 4.82  \\
    KESPEECH\_TEST           & 10.15  & 11.37  & 4.53  & 10.68  & 3.60  & 6.46  & 5.24  \\
    MD\_CONV\_DEV  & 8.08  & 7.81  & 4.62 & 9.54  & 5.10  & 7.02  & 25.69  \\
    MD\_CONV\_TEST & 10.70  & 10.56  & 6.36  & 12.09  & 6.92  & 9.45  & 36.48  \\
    MD\_READ\_DEV     & 4.36  & 4.12  & 0.79  & 5.09  & 1.41  & 2.71  & 1.69  \\
    MD\_READ\_TEST    & 4.02  & 4.00  & 0.92  & 4.29  & 1.46  & 2.71  & 1.73  \\
    SEAME\_DEV\_MAN          & 27.09  & 33.11  & 31.21 & 37.52 & 31.14  & 31.95  & 35.12  \\
    SEAME\_DEV\_SEG          & 39.22  & 53.73  & 50.98  & 78.76  & 51.75  & 53.80  & 54.42  \\
    SIO\_ASR\_ZH00000     & 2.82 & 2.58 & 2.28 & 3.30 & 2.30 & 2.64 & 2.36 \\
    SIO\_ASR\_ZH00001     & 1.04 & 0.61 & 0.79 & 1.56 & 0.59 & 0.70 & 0.52 \\
    SIO\_ASR\_ZH00002     & 4.44 & 3.51 & 3.00 & 5.02 & 2.88 & 3.50 & 3.71 \\
    SIO\_ASR\_ZH00003     & 2.45 & 1.19 & 1.13 & 3.19 & 0.95 & 1.00 & 0.94 \\
    SIO\_ASR\_ZH00004     & 2.22 & 1.77 & 1.57 & 2.75 & 1.59 & 2.09 & 1.61 \\
    SIO\_ASR\_ZH00005     & 2.79 & 2.16 & 2.24 & 3.72 & 2.12 & 2.62 & 1.91 \\
    SIO\_ASR\_ZH00006     & 6.33 & 5.26 & 4.81 & 7.39 & 4.79 & 6.26 & 5.46 \\
    SIO\_ASR\_ZH00007     & 6.71 & 4.95 & 3.67 & 10.23 & 3.78 & 7.44 & 5.42 \\
    SIO\_ASR\_ZH00008     & 5.33 & 4.34 & 4.10 & 8.50 & 4.00 & 6.65 & 5.08 \\
    SIO\_ASR\_ZH00009     & 4.06 & 3.41 & 3.54 & 4.98 & 3.31 & 3.67 & 3.38 \\
    SIO\_ASR\_ZH00010     & 3.87 & 3.43 & 3.36 & 4.32 & 3.31 & 3.50 & 3.53 \\
    SIO\_ASR\_ZH00011     & 2.02 & 1.51 & 1.37 & 3.18 & 1.40 & 1.56 & 1.33 \\
    SIO\_ASR\_ZH00012     & 3.64 & 3.04 & 2.27 & 4.49 & 2.06 & 3.24 & 2.30 \\
    SIO\_ASR\_ZH00013     & 4.17 & 3.53 & 4.28 & 5.95 & 4.00 & 3.79 & 4.08 \\
    SIO\_ASR\_ZH00014     & 5.08 & 4.21 & 3.53 & 8.41 & 3.55 & 4.20 & 3.77 \\
    SIO\_ASR\_ZH00015     & 7.33 & 5.21 & 8.16 & 12.09 & 7.00 & 6.47 & 5.73 \\
    SIO\_ASR\_ZH00016     & 6.83 & 5.43 & 5.49 & 9.14 & 5.22 & 5.64 & 5.15 \\
    SIO\_ASR\_ZH00017     & 3.75 & 2.76 & 2.65 & 5.53 & 2.47 & 2.94 & 2.56 \\
    SIO\_ASR\_ZH00018     & 3.39 & 3.14 & 2.54 & 4.25 & 2.44 & 3.45 & 3.04 \\
    SIO\_ASR\_ZH00019     & 4.32 & 3.84 & 3.43 & 7.11 & 3.31 & 4.10 & 3.30 \\
    SIO\_ASR\_ZH00020     & 2.59 & 1.32 & 1.63 & 3.76 & 1.34 & 1.58 & 1.34 \\
    SIO\_ASR\_ZH00021     & 3.73 & 3.10 & 2.81 & 5.23 & 2.72 & 3.26 & 2.65 \\
    SIO\_ASR\_ZH00022     & 5.87 & 5.07 & 4.12 & 7.02 & 3.52 & 4.63 & 3.39 \\
    SIO\_ASR\_ZH00023     & 3.26 & 2.80 & 2.48 & 4.45 & 2.18 & 2.59 & 2.78 \\
    SIO\_ASR\_ZH00024     & 6.43 & 4.97 & 4.95 & 10.08 & 4.55 & 5.90 & 4.99 \\
    SIO\_ASR\_ZH00025     & 5.05 & 4.49 & 3.87 & 6.42 & 3.65 & 4.53 & 4.37 \\
    SIO\_ASR\_ZH00026     & 4.79 & 4.14 & 4.32 & 5.51 & 3.99 & 4.62 & 3.57 \\
    THCHS-30\_DEV            & 4.72 & 3.76 & 0.09 & 5.05 & 0.32  & 2.71  & 1.10  \\
    THCHS-30\_TEST           & 5.18 & 3.98 & 0.27 & 5.67 & 0.56  & 3.07  & 1.36  \\
    WS\_DEV         & 3.49 & 3.14 & 3.21 & 7.53 & 3.23  & 4.72  & 3.11  \\
    WS\_TEST\_MEETING & 7.34  & 6.98  & 4.76  & 7.83 & 4.63  & 7.64  & 6.23  \\
    WS\_TEST\_NET   & 7.13  & 6.63  & 4.85 & 9.30 & 4.60  & 5.97  & 6.44  \\
    \midrule
    Overall                  & 8.46  & 8.18  & 5.66 & 10.86 & 5.68  & 8.03  & 9.84  \\
    \bottomrule
  \end{tabularx}
  }
  \caption{
    The WER (\%) results of Conventional ASR and Large Audio Languages Models on existing open-source Mandarin ASR benchmarks. The ``Overall'' results represent the average WER of each model on test speeches across all benchmarks. ``SV-Small" refers to SenseVoice-Small, ``PF-Large" to Paraformer-Large, and ``FASR" to FireredASR. ``AS" refers to AISHELL, ``AM" to AliMeeting, ``CV" to CommonVoice, ``MD" to MagicData, ``SIO" to SpeechIO, and ``WS" to WenetSpeech.
  }
  \label{tab:asr_zh_wer}
\end{table*}

\section{C. Prompts Using in Entity-rich Corpora Generation}
In this work, we utilize DeepSeek-R1 to generate entity-rich text corpora for ContextASR-Bench. 
For its two subsets, ContextASR-Speech and ContextASR-Dialogue, we craft two distinct prompts tailored to the specific characteristics of each subset. 
For ContextASR-Speech, we feed the LLM with open-source NER datasets from various domains, ensuring broad domain coverage for both the generated corpus and the named entities it contains. 
For ContextASR-Dialogue, we instead provide movie-related information—such as title, crew, and movie synopsis, prompting the model to produce dialogue corpora densely populated with named entities, especially personal names. 
The detailed prompts are presented below.
\label{appendix:prompt_for_corpora}

\subsection{C1. LLM Prompt For ContextASR-Speech}
\label{appendix:speech_prompt}
\begin{PromptBlock}
\textbf{\# Task Overview} \\
Convert formal written text for a Named Entity Recognition (NER) task into a more natural, spoken language style (as if spoken by a real person). Simultaneously, output the list of named entities appearing in the spoken text and the domain label of the generated text.

\medskip
\textbf{\# Input Details} \\
1. \textbf{Original Text}: A snippet of formal written text intended for a Named Entity Recognition task.  \\
2. \textbf{Entity List}: Named entities present in the original text (separated by semicolons ;). 

\medskip
\textbf{\# Output Details} \\
1. \textbf{Spoken Text}: A segment of text converted into a natural spoken style based on the provided formal written text. \\
2. \textbf{Entity List}: Named entities appearing in the generated spoken text (separated by semicolons ;).  \\
3. \textbf{Domain Label}: A simple word or short phrase indicating the domain of the generated spoken text and its entities, based on judgment.

\medskip
\textbf{\# Specific Requirements} \\
1. \textbf{Creativity \& Naturalness}: Use your imagination to generate natural-sounding spoken text, as if spoken by a real person. This could take the form of dialogue, a monologue, casual conversation, anecdote, etc. The text must be between 100 and 400 words in length.  \\
2. \textbf{Meaning \& Domain Relevance}: The generated spoken text does not need to convey the exact same meaning as the original written text. It only needs to belong to the same general domain. The original text serves as a guideline for the domain. \\
3. \textbf{Entity Usage Flexibility}: The generated spoken text does not need to include all named entities provided in the input list. The input entities are provided only for inspiration and domain context. \\
4. \textbf{Rich Entity Inclusion}: The spoken text must contain at least five named entities. Entities are not limited to standard types like person, location, or organization names. Prioritize including domain-specific terms, technical jargon, or specialized concepts relevant to the inferred domain. \\
5. \textbf{Text Normalization}: The generated spoken text must be normalized for readability, mimicking speech transcribed by an ASR tool. Use only common punctuation marks alongside English words and characters. Convert numbers, dates, mathematical units, currency symbols, etc., into their corresponding spoken English words. Examples: 2023-09-28 $\rightarrow$ September twenty-eighth, twenty twenty-three or the twenty-eighth of September, twenty twenty-three; 32.5 g/L $\rightarrow$ thirty-two point five grams per liter; 1000 \$ $\rightarrow$ one thousand dollars or a thousand dollars; 45\% $\rightarrow$ forty-five percent. \\
6. \textbf{Avoid Structured Text \& Emojis}: Do not include any structured text formats (e.g., Markdown, LaTeX, HTML, XML, JSON) or emojis in the generated spoken text. 

\medskip
\textbf{\# Examples} \\
\textbf{\#\#  Example 1} \\
\textbf{Original Text}: Serial anteroposterior and lateral chest radiographs (daily for the first 3-4 days). \\
\textbf{Entity List}: anteroposterior and lateral chest radiographs \\
Generated spoken text, entity list, and domain label: \\
\textbf{Spoken Text}: Hey Mary, I took my dad for his follow-up yesterday. The doctor ordered a chest X-ray, both AP and lateral views. They want him to get it done daily for like, the next three or four mornings? Seems like a lot of trips. Dr. Chen also mentioned possibly needing a CT scan later to get a clearer look at the pulmonary vasculature or any nodular lesions. Though Mr. Johnson in the next bed said all he had was an MRI. All these imaging options – radiography, CT, MRI – it's quite something nowadays. \\
\textbf{Entity List}: chest X-ray; AP and lateral views; CT scan; pulmonary vasculature; nodular lesions; radiography; CT; MRI \\
\textbf{Domain Label}: Healthcare 

\medskip
\textbf{\#\# Example 2} \\
\textbf{Original Text}: The Sishui County government office initiated a ban on unauthorized bian stone excavation in March. Many villages have established patrol teams to immediately halt any observed illicit digging. \\
\textbf{Entity List}: Sishui County government office; patrol teams \\
Generated spoken text, entity list, and domain label: \\
\textbf{Spoken Text}:Whoa, Bob, did you see the new county directive? Came down in March from the Sishui County Administration – totally banned any private digging for bian stone now. They've got inspection units out checking villages. Apparently, young Zhang got caught yesterday up near the ridge with his metal detector prospecting for mineral seams. Took just two swings with his pick before the conservation crew stopped him. Honestly, protecting that basalt formation makes sense. Remember that ground subsidence last month in Oak Valley from all the quarrying? Heard the Environmental Protection Agency might even issue portable spectrometers to help verify mineral composition on-site. \\
\textbf{Entity List}: Sishui County Administration; inspection units; bian stone; metal detector; mineral seams; conservation crew; basalt formation; ground subsidence; quarrying; Environmental Protection Agency; portable spectrometers; mineral composition \\
\textbf{Domain Label}: Natural Resource Management 

\medskip
\textbf{\# Now, please generate the spoken text, entity list, and domain label based on the above requirements and examples, using the provided original text and entity list as context.} 

\medskip
\textbf{Original Text}: \{raw\_text\} \\
\textbf{Entity List}: \{entities\} \\
Generated spoken text, entity list, and domain label:
\end{PromptBlock}

\subsection{C2. LLM Prompt For ContextASR-Dialogue}
\label{appendix:dialogue_prompt}
\begin{PromptBlock}

\textbf{\# Task Overview} \\
Generate a natural multi-person conversation script in English about a specific movie, along with an entity list, based on provided information.

\medskip
\textbf{\# Input Details} \\
1. \textbf{Movie Title}: Title of the film being discussed. \\
2. \textbf{Director}: Name of the film's director. \\
3. \textbf{Cast}: Main actors/actresses in the film. \\
4. \textbf{Plot Summary}: Brief synopsis of the movie's storyline. \\
5. \textbf{Number of Participants}: Total people in the conversation.

\medskip
\textbf{\# Output Requirements} \\
1. \textbf{Dialogue Script}: Casual conversation in screenplay format (``Speaker: Dialogue''). \\
2. \textbf{Entity List}: Proper nouns (titles, names) and film-related terminology from the conversation (semicolon-separated).

\medskip
\textbf{\# Key Specifications} \\
1. \textbf{Participant Names}: Use culturally appropriate Western names (e.g., Chris, Emily, Marcus, Rachel) matching participant count. \\
2. \textbf{Natural Dialogue}: The content of the conversation must be highly colloquial, natural and fluent, in line with the true context of easy discussion of the movie between friends or fans, avoiding any written language or blunt wording, and should be full of life and personal opinions. \\
3. \textbf{Movie Content Focus}: The core content of the discussion must revolve around the provided movie, engaging in an in-depth analysis. If the provided plot summary is not detailed enough, please reasonably supplement and expand by incorporating information you are aware of regarding the movie (such as a more detailed plot, background, themes, reviews, etc.) to enrich the conversation content, making it more profound and comprehensive. You may include perspectives on the plot, characters, actors' performances, directorial techniques, thematic significance, and other aspects. \\
4. \textbf{Entity Integration}: Mention at least 3 movie-related names (director/actor/character), with at least 2 names naturally mentioned by different speakers. Please pay special attention that all names mentioned in the dialogue must be in their full form. For example, use ``Robin White'' instead of just ``White''. \\
5. \textbf{Dialogue Coherence}: The entire generated dialogue text should possess a high degree of realism and logical coherence, ensuring that statements between different speakers naturally follow, respond to, and advance the discussion topic, forming an organic and complete dialogue process. \\
6. \textbf{Text Normalization}: The dialogue text needs to undergo ``normalization'' processing to simulate speech transcription effects. All numbers (such as years, times, quantities, rankings, dates, currencies, units, etc.) should be converted into their corresponding English words (for example, ``nineties'' instead of ``90s'', ``eight-thirty'' not ``8:30'', ``three hundred dollars'' not ``300\$'') instead of using Arabic numerals or symbols. \\
7. \textbf{Entity Extraction}: Carefully review the generated dialogue text, extract all movie titles, directors, actors, main character names, as well as professional terms related to movie production and film reviews mentioned in the dialogue text, and organize them into a list separated by semicolons ``;''. Please ensure that all entities in the list accurately appear in the generated dialogue text. \\
8. \textbf{Format Rules}: It is strictly prohibited to use any structured markup languages (such as Markdown, LaTeX, HTML, XML, JSON, etc.) or emoticons in the generated dialogue text. The dialogue content should only include English words or letters, and common punctuation marks, with each line presented in the format ``Speaker: Content''. Each speaker in the entire dialogue must have no fewer than three utterances.

\medskip
\textbf{\# Language Specific Instructions} \\
The provided movie information is in Chinese (including movie titles, directors' and cast members' full names, and plot summary). When generating English dialogue based on provided Chinese movie information ensure the following: \\
1. \textbf{Movie Title}: Do not simply translate the Chinese titles. Instead, use the official English titles of the movies. \\
2. \textbf{Personal Names}: Use the authentic English names of directors, cast members and characters instead of directly transliterating from Chinese. \\
3. \textbf{Plot Summary}: While translating the plot summary, maintain the original context and nuance without altering the intended meaning. Ensure that cultural references are appropriately adapted for an English-speaking audience.

\medskip
\textbf{\# Now, please generate dialogue script and entity list based on the above requirements and the provided movie information, referring to the example's reply format.} 

\medskip
\textbf{Movie Title}: \{movie\_name\} \\ 
\textbf{Director}: \{movie\_director\} \\
\textbf{Cast}: \{movie\_actors\} \\
\textbf{Plot Summary}: \{movie\_plot\} \\
\textbf{Number of Participants}: \{person\_num\} \\
Generated dialogue script and entity list:
\end{PromptBlock}

\section{D. User Prompts for Contextual Speech Recognition}
\label{appendix:infer_prompt}
In Section 3, we conduct a comprehensive evaluation of existing open-source LALMs on ContextASR-Bench. Below, we list the exact user prompts used for each model, enabling researchers to reproduce our experimental results precisely. 

\subsection{D1. User Prompts for ContextASR-Speech}
\textbf{Contextless Setting}: 
\begin{itemize}
\item \textbf{Qwen2-Audio}: ``Detect the language and recognize the speech:"
\item \textbf{Kimi-Audio}: ``Please transcribe the following audio:"
\item \textbf{Baichuan-Audio}: ``Transcribe the speech into text:"
\item \textbf{Baichuan-Omni-1.5}: ``Transcribe the speech into text:"
\item \textbf{Qwen2.5-Omni}: ``Transcribe the English audio into text, ensuring all punctuation marks are included."
\end{itemize}

\textbf{Coarse-grained Context Setting}: 
\begin{itemize}
\item \textbf{Qwen2-Audio}: ``This speech belongs to the \textless domain label\textgreater~ field. Detect the language and recognize the speech:"
\item \textbf{Kimi-Audio}: ``The following audio belongs to the \textless domain label\textgreater~ field. Please transcribe the following audio:"
\item \textbf{Baichuan-Audio}: ``This speech belongs to the \textless domain label\textgreater~ domain. Transcribe the speech into text:"
\item \textbf{Baichuan-Omni-1.5}: ``This speech belongs to the \textless domain label\textgreater~ domain. Transcribe the speech into text:"
\item \textbf{Qwen2.5-Omni}: ``This audio belongs to the \textless domain label\textgreater~ field. Transcribe the English audio into text, ensuring all punctuation marks are included."
\end{itemize}

\textbf{Fine-grained Context Setting}: 
\begin{itemize}
\item \textbf{Qwen2-Audio}: ``This speech belongs to the \textless domain label\textgreater~ field and may contains
the following words or phrases: \textless entity list\textgreater. Detect the language and recognize the speech:"
\item \textbf{Kimi-Audio}: ``The following audio belongs to the \textless domain label\textgreater~ field and may
contains the following words or phrases: \textless entity list\textgreater. Please transcribe the following audio:"
\item \textbf{Baichuan-Audio}: ``This speech belongs to the \textless domain label\textgreater~ domain and may contain the following words or phrases: \textless entity list\textgreater. Transcribe the speech into text:"
\item \textbf{Baichuan-Omni-1.5}: ``This speech belongs to the \textless domain label\textgreater~ domain and may contain the following words or phrases: \textless entity list\textgreater. Transcribe the speech into text:"
\item \textbf{Qwen2.5-Omni}: ``This audio belongs to the \textless domain label\textgreater~ field and may contains the following words or phrases: \textless entity list\textgreater. Transcribe the English audio into text, ensuring all punctuation marks are included."
\end{itemize}

\subsection{D2. User Prompts for ContextASR-Dialogue}
\textbf{Contextless Setting}: 
\begin{itemize}
\item \textbf{Qwen2-Audio}: ``Detect the language and recognize the speech:"
\item \textbf{Kimi-Audio}: ``Please transcribe the following audio:"
\item \textbf{Baichuan-Audio}: ``Transcribe the speech into text:"
\item \textbf{Baichuan-Omni-1.5}: ``Transcribe the speech into text:"
\item \textbf{Qwen2.5-Omni}: ``Transcribe the English audio into text, ensuring all punctuation marks are included."
\end{itemize}

\textbf{Coarse-grained Context Setting}: 
\begin{itemize}
\item \textbf{Qwen2-Audio}: ``This speech is a dialogue about the movie \textless movie title\textgreater. Detect the language and recognize the speech:"
\item \textbf{Kimi-Audio}: ``The following audio is a dialogue about the movie \textless movie title\textgreater. Please transcribe the following audio:"
\item \textbf{Baichuan-Audio}: ``This speech is a dialogue about the movie \textless movie title\textgreater. Transcribe the speech into text:"
\item \textbf{Baichuan-Omni-1.5}: ``This speech is a dialogue about the movie \textless movie title\textgreater.  Transcribe the speech into text:"
\item \textbf{Qwen2.5-Omni}: ``This audio is a dialogue about the movie \textless movie title\textgreater.  Transcribe the English audio into text, ensuring all punctuation marks are included."
\end{itemize}

\textbf{Fine-grained Context Setting}: 
\begin{itemize}
\item \textbf{Qwen2-Audio}: ``This speech is a dialogue about the movie \textless movie title\textgreater~ and may contains the following words or phrases: \textless entity list\textgreater. Detect the language and recognize the speech:"
\item \textbf{Kimi-Audio}: ``The following audio is a dialogue about the movie \textless movie title\textgreater~ and may contains the following words or phrases: \textless entity list\textgreater. Please transcribe the following audio:"
\item \textbf{Baichuan-Audio}: ``The speech is a dialogue about the movie \textless movie title\textgreater, and may contain the following words or phrases: \textless entity list\textgreater. Transcribe the speech into text:"
\item \textbf{Baichuan-Omni-1.5}: ``The speech is a dialogue about the movie \textless movie title\textgreater, and may contain the following words or phrases: \textless entity list\textgreater. Transcribe the speech into text:"
\item \textbf{Qwen2.5-Omni}: ``This audio is a dialogue about the movie \textless movie title\textgreater~ and may contains the following words or phrases: \textless entity list\textgreater. Transcribe the English audio into text, ensuring all punctuation marks are included."
\end{itemize}

\end{document}